\documentclass[aps,prb,twocolumn,showpacs,groupedaddress]{revtex4}

\usepackage{graphicx}
\usepackage{latexsym}
\usepackage{amssymb}

\begin{document}

\title{Ordered spin structures of $\beta$-MnO$_{2}$ (rutile-type) systems with competing exchange interactions: numerical approach using equi-energy contour plot}

\author{Masatsugu Suzuki }
\email[]{suzuki@binghamton.edu}
%\thanks{}
\affiliation{Department of Physics, State University of New York at Binghamton, Binghamton, New York 13902-6000}

\author{Itsuko S. Suzuki }
\email[]{itsuko@binghamton.edu}
%\thanks{}
\affiliation{Department of Physics, State University of New York at Binghamton, Binghamton, New York 13902-6000}


\date{\today}

\begin{abstract}
Using numerical calculations of equi-energy contour plot of the Fourier transform of the spin Hamiltonian, we study the magnetic phase diagram ($J_{2}$ and $J_{3}$) of the rutile type $\beta$-MnO$_{2}$, where $J_{1}$ ($<0$) is fixed and is the antiferromagnetic interaction along the diagonal direction, $J_{2}$ is the interaction along the $c$ axis, and $J_{3}$ is the interaction along the $a$ axis. The magnetic phase diagram consists of the multricritical point (the intersection $J_{2}J_{3}=J_{1}^{2}$ and $J_{2}+J_{3}=2J_{1}$), the helical order along the $c$ axis, the $(h=1/2,k=0,l=1/2)$ phase, the helical order along the $a$ axis, and the phase $(h=0,k=0,l=1)$. The shift of the location of the magnetic Bragg peak in the $(h,0,l)$ reciprocal lattice plane is examined with the change of $J_{2}$ and $J_{3}$ in the phase diagram. The shift is discontinuous on the first-order phase transition, and is continuous on the second-order phase transition. The detail of our magnetic phase diagram is rather different from that reported by Yoshimori.
\end{abstract}

%\keywords{}

\maketitle

%\section{}
%\subsection{}
%\subsubsection{}

% If in two-column mode, this environment will change to single-column
% format so that long equations can be displayed. Use
%\begin{widetext}
%\end{widetext}

\section{\label{intro}Introduction}
It is well known that $\beta$-MnO$_{2}$ (with the rutile-type structure) is one of the systems with a helical spin order along the $c$ axis. Erickson\cite{ref01} found from magnetic neutron scattering on $\beta$-MnO$_{2}$ that the spins are helically ordered with a period of 7$c$ along the $c$ axis. The direction of spins in the $c$ plane turns from one plane to the next plane by an angle of $129^{\circ}$ ($5\pi/7$). In 1958, Yoshimori\cite{ref02} theoretically demonstrated the origin and stability of the helical spin order in $\beta$-MnO$_{2}$, based on the experimental results from Erickson. The spin Hamiltonian of $\beta$-MnO$_{2}$ is described by a form of the Heisenberg-type, where the combination of nearest neighbor (n.n.) exchange interactions, next nearest neighbor (n.n.n.) exchange interactions, and so on are included. His success lies in the use of the Fourier transform of the spin Hamiltonian in the reciprocal lattice space. Then the Fourier transform depends only on exchange interactions and wavevectors {\bf Q}. The location of the magnetic Bragg points in the reciprocal lattice space are uniquely determined from the condition that the Fourier transform takes a maximum. 

In spite of the success in theory established by Yoshimori,\cite{ref02} Erickson has not published his detailed data of the magnetic neutron scattering of $\beta$-MnO$_{2}$ as far as we know. Since 1958, there have been several papers on the magnetic structure of $\beta$-MnO$_{2}$ using magnetic neutron scattering\cite{ref03,ref04,ref05} and magnetic x-ray scattering.\cite{ref06,ref07} In recent years, Sato et al, have reported that the magnetic Bragg peaks appear at the wave vector {\bf Q} = (1, 0, 2 + $\epsilon$) in the units of $a^{*}$ ($=2\pi/a$), $b^{*}$ ($=2\pi/b$), and $c^{*}$ ($=2\pi/c$) in the reciprocal lattice space, where $\epsilon$ = 0.297 at 10 K, increases with increasing temperature, and reaches 0.2992 just below $T_{N}$ (= 92 K). The value of $\epsilon$ is rather different from 2/7 derived by Yoshimori.\cite{ref02} This indicates that the helical spin structure along the $c$ axis is incommensurate with the $c$- axis lattice constant. 

Because of the crystal field (distorted octahedron formed by O$^{2-}$ ions) in the vicinity of Mn$^{4+}$ ion, the ground orbital state of Mn$^{4+}$ ion ($3d^{3}$, $L$ = 3 and $S$ = 3/2) is split into the $t_{2g}$ ($d\epsilon$) level (lower energy level, triple degenerate) and the $e_{g}$ ($d\gamma$) level (upper energy, double degenerate). As a result, the ground state is now orbital singlet, indicating that the orbital angular momentum is quenched. The $e_{g}$ electrons are responsible for the metallic conduction, while the localized $t_{2g}$ electrons are responsible for the magnetism. In recent years, Sato et al.\cite{ref08} have reported the transport properties of a single crystal $\beta$-MnO$_{2}$, such as electrical resistivity, thermopower, Hall effect, and magnetoresistance. Their data show an appreciable anomaly near $T_{N}$. This implies that there is a strong correlation between conduction $\epsilon_{g}$ electrons and localized $t_{2g}$ magnetic moment through the Hund's rule. The DC magnetic susceptibility shows a significant deviation from a molecular-field theory based on a localized spin model, which was assumed by Yoshimori\cite{ref02} on his helical spin order. 

In the present paper, we study the magnetic phase diagram of $\beta$-MnO$_{2}$ type structure using the molecular field theory developed by Yoshimori,\cite{ref02} where Mn$^{4+}$ spins are localized. The exchange interactions $J_{1}$ along the diagonal axis of the system, $J_{2}$ along the $c$ axis and $J_{3}$ along the $a$ axis are taken into account. Note that $J_{1}$ is assumed to be fixed and be antiferromagnetic. According to Sato et al.,\cite{ref08} the intra-atomic $t_{2g}-e_{g}$ exchange interaction (Hund coupling), the transfer interaction between adjacent $e_{g}$ orbitals, and the occupancy of the $e_{g}$ orbitals are defined by $J_{Hund}$, $t$, and $c$, respectively. Since $\left| J_{n}\right|\gg \left| cJ_{Hund}\right|$ and $\left| J_{n}\right|\gg\left| ct\right|$ for $\beta$-MnO$_{2}$ ($n=1,2,3$),\cite{ref08} this means that the spin structure of $\beta$-MnO$_{2}$ below $T_{N}$ is not affected by the effect of $cJ_{Hund}$ and $ct$ at al. In other words, the spin structure is well described by the localized spin model with $J_{1}$, $J_{2}$, and $J_{3}$. We find that our magnetic phase diagram ($J_{2}$ vs $J_{3}$) consists of four phases including the helical phase along the $c$ axis, the phase with (1/2, 0, 1/2), the helical phase along the $a$ axis, and the phase with (1, 0, 0). The detail of our phase diagram is rather different from that proposed by Yoshimori.\cite{ref02} We use a numerical calculation approach in finding the distribution of the magnetic Bragg peaks in the fixed reciprocal lattice planes such as $(h,k,l)$ with one index fixed. To this end we calculate the equi-energy contour plot of the negative sign of the Fourier transform of the spin Hamiltonian, $J(h,k,l)$. The magnetic Bragg peaks are located inside the maximum equi-energy contour. The selection rule for the location of the magnetic Bragg peaks is the same as that derived by Yoshimori.\cite{ref02} This numerical method has an advantage in visualizing the location of the magnetic Bragg peaks in the reciprocal lattice space. The nature of the phase transitions on the phase boundaries will be discussed. 

\section{\label{back}BACKGROUND: general theory for the ordered spin structure}
We follow the theory presented by Nagamiya.\cite{ref09} We consider a lattice of magnetic atoms such as $\beta$-MnO$_{2}$. The unit cell can be chosen so that it contains one magnetic atom. On each magnetic atom, we assume a \textit{classical spin}. Between the spin ${\bf S}_{i}$ at the position ${\bf R}_{i}$ and ${\bf S}_{j}$ at ${\bf R}_{j}$, there is an Heisenberg-type exchange interaction. The Heisenberg spin Hamiltonian is expressed by 
\begin{equation}
H=-2\sum\limits_{i,j}J({\bf R}_{ij} ){\bf S}_{i} \cdot {\bf S}_{j} ,
\label{eq01}
\end{equation} 
where
\[ 
J(-{\bf R}_{ij})=J({\bf R}_{ij}) ,
\] 
and
\[ 
{\bf R}_{ij} ={\bf R}_{i} -{\bf R}_{j} .
\] 
The exchange interaction $J({\bf R}_{ij})$ is not restricted to the nearest neighbors. We now use the Fourier transformations of the exchange interaction and spin;
\begin{eqnarray} 
J({\bf q})=\sum\limits_{j(\neq i)}J({\bf R}_{ij} )\exp (-i{\bf q}\cdot {\bf R}_{ij} ) , \\
\label{eq02} 
{\bf S}_{i} =\frac{1}{\sqrt{N} } \sum\limits_{{\bf q}}{\bf S}_{{\bf q}} \exp (iq\cdot {\bf R}_{i} ) ,
\label{eq03}
\end{eqnarray} 
with
\[ 
{\bf S}_{{\bf q}} =\frac{1}{\sqrt{N} } \sum\limits_{i}{\bf S}_{i} \exp (-i{\bf q}\cdot {\bf R}_{i} )  ,
\] 
where $N$ ($= N_{1}N_{2}N_{3}$) is the total number of spins, and ${\bf S}_{q}^{*}={\bf S}_{-q}$. The position vector ${\bf R}_{i}$ is expressed by 
\[ 
{\bf R}_{i}=n_{1}{\bf a}_{1}+n_{2}{\bf a}_{2}+n_{3}{\bf a}_{3}\text{  (}n_{1}, n_{2}, n_{3}\text{ are integers)} ,
\]
where $n_{1}=0,1,\cdots,N_{1}$, $n_{2}=0,1,\cdots,N_{2}$, $n_{3}=0,1,\cdots,N_{3}$, and ${\bf a}_{1}$, ${\bf a}_{2}$, and ${\bf a}_{3}$ are the fundamental lattice vectors. We also define the reciprocal lattice vector $G$ by 
\[
G(h,k,l)=h{\bf b}_{1} +k{\bf b}_{2} +l{\bf b}_{3} \text{  (}h, k, l\text{ are integers)} ,
\] 
where {\bf b}$_{1}$, {\bf b}$_{2}$, and {\bf b}$_{3}$ are fundamental reciprocal lattice vectors and are given by 
\[
{\bf b}_{1} =2\pi \frac{{\bf a}_{2} \times {\bf a}_{3} }{\lbrack {\bf a}_{1} ,{\bf a}_{2} ,{\bf a}_{3} \rbrack} , 
{\bf b}_{2} =2\pi \frac{{\bf a}_{3} \times {\bf a}_{1} }{\lbrack {\bf a}_{1} ,{\bf a}_{2} ,{\bf a}_{3} \rbrack} ,
{\bf b}_{3} =2\pi \frac{{\bf a}_{1} \times {\bf a}_{2} }{\lbrack {\bf a}_{1} ,{\bf a}_{2} ,{\bf a}_{3} \rbrack} ,
\]
with
\[ 
\lbrack {\bf a}_{1},{\bf a}_{2},{\bf a}_{3}\rbrack={\bf a}_{1}\cdot \left({\bf a}_{2}\times {\bf a}_{3}\right)={\bf a}_{2}\cdot\left({\bf a}_{3}\times {\bf a}_{1}\right)={\bf a}_{3} \cdot\left( {\bf a}_{1}\times {\bf a}_{2}\right).
\]
Noting that
\[ 
{\bf a}_{1} \cdot {\bf b}_{1} =2\pi  , 
{\bf a}_{2} \cdot {\bf b}_{2} =2\pi  , 
{\bf a}_{3} \cdot {\bf b}_{3} =2\pi  ,
\]
we have
\[ 
{\bf G}(h,k,l)\cdot {\bf R}_{i} =2\pi (n_{1}h+n_{2}k+n_{3}l)=2\pi\times \text{integer} . 
\]
The periodic boundary condition for ${\bf S}_{i}$ leads to 
\[
\exp \lbrack i{\bf q}\cdot (N_{1} {\bf a}_{1})\rbrack=1 , 
\exp \lbrack i{\bf q}\cdot (N_{2} {\bf a}_{2})\rbrack=1 , 
\exp \lbrack i{\bf q}\cdot (N_{3} {\bf a}_{3})\rbrack=1 . 
\]
This means that the wavevector {\bf q} is given by
\[ 
{\bf q}=q_{1} {\bf b}_{1} +q_{2} {\bf b}_{2} +q_{3} {\bf b}_{3} , 
\]
where
\[ 
q_{1} =\frac{m_{1} }{N_{1} }  , 
q_{2} =\frac{m_{2} }{N_{2} }  , 
q_{3} =\frac{m_{3} }{N_{2} }  . 
\]
For convenience we assume that
\[ 
-\frac{N_{1} }{2} \leq m_{1} \leq \frac{N_{1} }{2}  , 
-\frac{N_{2} }{2} \leq m_{2} \leq \frac{N_{2} }{2}  , 
-\frac{N_{3} }{2} \leq m_{3} \leq \frac{N_{3} }{2}  ,
\]
corresponding to the first Brillouin zone. There are $N_{1}N_{2}N_{3}=N$ wavevectors in the first Brillouin zone. The spin Hamiltonian is rewritten as 
\begin{equation} 
H=-\sum\limits_{{\bf q}}J({\bf q}){\bf S}_{{\bf q}} \cdot {\bf S}_{-{\bf q}}  .
\label{eq04}
\end{equation} 
We look for the lowest minimum of Eq.(\ref{eq01}) under the condition that 
\begin{equation} 
{\bf S}_{i}^{2}=S^{2}=\frac{1}{N} \sum\limits_{{\bf q},{\bf q}^{\prime}}{\bf S}_{{\bf q}} \cdot {\bf S}_{-{\bf q}^{\prime}} \exp \lbrack i({\bf q}-{\bf q}^{\prime})\cdot {\bf R}_{i} \rbrack ,
\label{eq05}
\end{equation} 
for any $i$. Instead of this condition, we impose a milder condition 
\begin{eqnarray} 
NS^{2} &=& \sum\limits_{i}{\bf S}_{i}^{2} 
=\frac{1}{N}\sum\limits_{i}\sum\limits_{{\bf q},{\bf q}^{\prime}}{\bf S}_{{\bf q}} \cdot {\bf S}_{-{\bf q}^{\prime}} \exp \lbrack i({\bf q}-{\bf q}^{\prime})\cdot R_{i} \rbrack  \nonumber \\
&=& \sum\limits_{{\bf q}}{\bf S}_{{\bf q}} \cdot {\bf S}_{-{\bf q}} ,
\label{eq06}
\end{eqnarray} 
where we use
\[ 
\sum\limits_{i}\exp \lbrack i({\bf q}-{\bf q}^{\prime})\cdot {\bf R}_{i} \rbrack =N\delta _{{\bf q},{\bf q}^{\prime}} .
\] 
Under this milder condition, the minimum of Eq.(\ref{eq01}) is obtained simply by taking only that {\bf q} for which $J({\bf q})$ has the maximum. Denoting this {\bf q} by {\bf Q} ({\bf q} = -{\bf Q} being equally allowed), we have the minimum value of Eq.(\ref{eq04}) as
\[ 
-J({\bf Q})({\bf S}_{{\bf Q}} \cdot {\bf S}_{-{\bf Q}} +{\bf S}_{-{\bf Q}} \cdot {\bf S}_{{\bf Q}} )  . 
\] 
We also obtain
\[ 
{\bf S}_{i} =\frac{1}{\sqrt{N}}\lbrack {\bf S}_{{\bf Q}}\exp (i{\bf Q}\cdot {\bf R}_{i})+{\bf S}_{-{\bf Q}}\exp (-i{\bf Q}\cdot {\bf R}_{i})\rbrack .
\] 
The condition (\ref{eq06}) can be written as
\begin{eqnarray*}
NS^{2}&=&2{\bf S}_{{\bf Q}} \cdot {\bf S}_{-{\bf Q}} +{\bf S}_{{\bf Q}} \cdot {\bf S}_{{\bf Q}} \exp \lbrack 2i{\bf Q}\cdot {\bf R}_{i} ] \\
&+&{\bf S}_{-{\bf Q}} \cdot {\bf S}_{-{\bf Q}} \exp \lbrack -2i{\bf Q}\cdot {\bf R}_{i} \rbrack , 
\end{eqnarray*}
for any ${\bf R}_{i}$. This indicates that
\begin{equation} 
{\bf S}_{{\bf Q}} \cdot {\bf S}_{{\bf Q}} =0 . 
\label{eq07}
\end{equation} 
Here we assume that
\begin{eqnarray} 
{\bf S}_{{\bf Q}} ={\bf R}_{{\bf Q}} +iI_{{\bf Q}} , \nonumber \\
{\bf S}_{-{\bf Q}} ={\bf R}_{{\bf Q}} -iI_{{\bf Q}} , \nonumber 
\end{eqnarray}
where {\bf R}$_{{\bf Q}}$ and ${\bf I}_{Q}$ are real vectors.
\begin{eqnarray} 
{\bf S}_{{\bf Q}} \cdot {\bf S}_{-{\bf Q}} &=& ({\bf R}_{{\bf Q}} +i{\bf I}_{{\bf Q}} )\cdot ({\bf R}_{{\bf Q}} -i{\bf I}_{{\bf Q}} ) \nonumber \\
&=&{\bf R}_{{\bf Q}} \cdot {\bf R}_{{\bf Q}} +{\bf I}_{{\bf Q}} \cdot {\bf I}_{{\bf Q}}  \nonumber \\
{\bf S}_{{\bf Q}} \cdot {\bf S}_{{\bf Q}} &=&({\bf R}_{{\bf Q}} +i{\bf I}_{{\bf Q}} )\cdot ({\bf R}_{{\bf Q}} +i{\bf I}_{{\bf Q}} ) \nonumber \\
&=& {\bf R}_{{\bf Q}} \cdot {\bf R}_{{\bf Q}} -{\bf I}_{{\bf Q}} \cdot I_{{\bf Q}} +2iR_{{\bf Q}} \cdot {\bf I}_{{\bf Q}} =0 .
\label{eq08}
\end{eqnarray}
Then we have
\begin{eqnarray} 
R_{{\bf Q}} =I_{{\bf Q}}  \nonumber \\
{\bf R}_{{\bf Q}} \cdot {\bf I}_{{\bf Q}} =0 \nonumber .
\end{eqnarray}
Then the vector ${\bf R}_{{\bf Q}}$ is perpendicular to the vector ${\bf I}_{{\bf Q}}$, and the magnitude $R_{{\bf Q}}$ is the same as the magnitude $I_{{\bf Q}}$. From Eq.(\ref{eq08}), we get 
\[ 
R_{{\bf Q}}^{2}=I_{{\bf Q}}^{2} =\frac{1}{4} NS^{2} . 
\] 
The minimum energy $E_{min}$ is obtained as
\begin{equation} 
E_{min}=-NS^{2}J({\bf Q})  .
\label{eq09}
\end{equation} 
The spin vector is expressed by
\begin{eqnarray*}
{\bf S}_{i}&=&\frac{1}{\sqrt{N}} [ ({\bf R}_{{\bf Q}}+i{\bf I}_{{\bf Q}})\exp (i{\bf Q}\cdot {\bf R}_{i})  \\
&+&({\bf R}_{{\bf Q}}-i{\bf I}_{{\bf Q}})\exp (-i{\bf Q}\cdot {\bf R}_{i})]  \\
&=&\frac{2}{\sqrt{N}}\lbrack ({\bf R}_{{\bf Q}}\cos ({\bf Q}\cdot {\bf R}_{i})-I_{{\bf Q}}\sin ({\bf Q}\cdot {\bf R}_{i})\rbrack \\
&=& S\lbrack ({\bf \hat{R}}_{{\bf Q}}\cos ({\bf Q}\cdot {\bf R}_{i})-{\bf \hat{I}}_{{\bf Q}}\sin ({\bf Q}\cdot {\bf R}_{i})\rbrack  ,
\end{eqnarray*}
where ${\bf \hat{R}}_{{\bf Q}}$ and ${\bf \hat{I}}_{{\bf Q}}$ are the unit vectors which are perpendicular to each other. For convenience, ${\bf \hat{R}}_{{\bf Q}}$ and ${\bf \hat{I}}_{{\bf Q}}$ are in the $x$-$y$ plane. The $z$ axis is perpendicular to the $x$-$y$ plane. Then using the unit vectors ${\bf e}_{x}$ and ${\bf e}_{y}$, we get the final result
\begin{equation} 
{\bf S}_{i}=S\lbrack\cos ({\bf Q}\cdot {\bf R}_{i}+\phi ){\bf e}_{x}+\sin ({\bf Q}\cdot {\bf R}_{i}+\phi ){\bf e}_{y}\rbrack ,
\label{eq10}
\end{equation} 
where $\phi$ is the angle between ${\bf \hat{R}}_{{\bf Q}}$ and $x$ axis. 

The spin structure thus derived is the most fundamental spin structure and is realized as a result of the minimum energy state in the classical spin system. The ferromagnetic state (${\bf Q} = 0$) and antiferromagnetic state ({\bf Q} = the zone boundary of the first Brillouin zone) are the special case of the spin structures. In general, {\bf Q} is not related to the crystal structure, but is related to the details of the exchange interactions. 

\section{\label{helical}Helical spin order for $\beta$-$\text{MnO}_{2}$}
\subsection{\label{helicalA}Calculation of $J({\bf q})$ for $\beta$-MnO$_{2}$} 

\begin{figure}
\includegraphics[width=7.0cm]{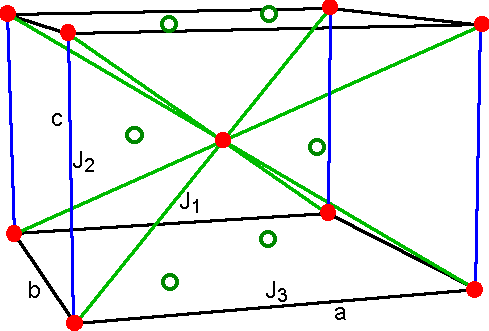}
\caption{\label{fig01}(Color online) The position of Mn$^{4+}$ ions (denoted by solid circles) and O$^{2-}$ (denoted by open circles) in $\beta$-MnO$_{2}$ (rutile-type structure, space group P4$_{2}$/mmm). $a=b=4.396\AA$. $c=2.871\AA$. $J_{1}$, $J_{2}$, and $J_{3}$ are the exchange interactions between Mn$^{2+}$ spins. One of the O atom is located at $(ua, ua)$, where $u=0.302$.}
\end{figure}

Figure \ref{fig01} shows the structure (rutile) of $\beta$-MnO$_{2}$ where $a=b=4.396\AA$, $c=2.871\AA$, and $u=0.302$.\cite{ref08} The exchange interactions $J_{1}$, $J_{2}$, and $J_{3}$ are defined by Fig.~\ref{fig01}. The definition of $J_{1}$, $J_{2}$, and $J_{3}$ are the same as that used by Yoshimori. The N\'{e}el temperature $T_{N}$ is equal to 92 K. Each Mn$^{4+}$ ion and surrounding six O$^{2-}$ ions form a cation-occupied deformed octahedron (see Sec.~\ref{dis}). 

Here we calculate $J({\bf q})$ for MnO$_{2}$, where the wavenumber {\bf q}, the lattice vectors ${\bf a}(n)$ ($n$ = 1, 2,\dots , 4) and {\bf c} are given by 
\begin{eqnarray*} 
{\bf q} &=& (q_{x}, q_{y}, q_{z}),  \\ 
{\bf a}(n)&=&{a\cos\lbrack\frac{\pi}{2}(n-1)\rbrack, a\sin\lbrack\frac{\pi}{2}(n-1)\rbrack, 0},  \\
{\bf c} &=& (0, 0, c). 
\end{eqnarray*} 
Then the expression of $J({\bf q})$ is given by
\begin{widetext}
\begin{eqnarray}
J({\bf q}) & = & J_{1} \exp \lbrack i{\bf q}\cdot\frac{{\bf c}+{\bf a}(1)+{\bf a}(2)}{2}\rbrack+J_{1} \exp \lbrack i{\bf q}\cdot\frac{-{\bf c}+{\bf a}(1)+{\bf a}(2)}{2} \rbrack \nonumber \\
&  & +J_{1} \exp \lbrack i{\bf q}\cdot\frac{{\bf c}+{\bf a}(2)+{\bf a}(3)}{2} \rbrack+J_{1} \exp \lbrack i{\bf q}\cdot\frac{-{\bf c}+{\bf a}(2)+{\bf a}(3)}{2} \rbrack \nonumber \\
&  & +J_{1} \exp \lbrack i{\bf q}\cdot\frac{{\bf c}+{\bf a}(3)+{\bf a}(4)}{2} \rbrack+J_{1} \exp \lbrack i{\bf q}\cdot\frac{-{\bf c}+{\bf a}(3)+{\bf a}(4)}{2} \rbrack \nonumber \\
&  & +J_{1} \exp \lbrack i{\bf q}\cdot\frac{{\bf c}+{\bf a}(4)+{\bf a}(1)}{2} \rbrack+J_{1} \exp \lbrack i{\bf q}\cdot\frac{-{\bf c}+{\bf a}(4)+{\bf a}(1)}{2} \rbrack \nonumber \\
&  & +J_{2} \exp \lbrack i{\bf q}\cdot {\bf c}\rbrack+J_{2} \exp \lbrack -i{\bf q}\cdot {\bf c}\rbrack \\
&  & +J_{3} \exp \lbrack i{\bf q}\cdot {\bf a}(1)\rbrack+J_{3} \exp \lbrack -i{\bf q}\cdot {\bf a}(1)\rbrack+J_{3} \exp \lbrack i{\bf q}\cdot {\bf a}(2)\rbrack+J_{3} \exp \lbrack -i{\bf q}\cdot {\bf a}(2)\rbrack  . \nonumber 
\end{eqnarray}
\end{widetext}
Then $J({\bf q})$ can be rewritten as
\begin{eqnarray} 
J({\bf q})&=&J(h,k,l)=8J_{1}\cos (\pi h)\cos (\pi k)\cos (\pi l) \nonumber \\
&+&2J_{2}\cos (2\pi l)+2J_{3}\lbrack\cos (2\pi h)+\cos (2\pi k)\rbrack , \nonumber \\
\label{eq11}
\end{eqnarray} 
where
\[ 
q_{x}= (\frac{2\pi}{a})h =a^{*}h, \text{  }  q_{y}=(\frac{2\pi}{a})k=a^{*}k,  \text{  } q_{z}=(\frac{2\pi}{c})l = c^{*}l
\]
with $h$, $k$, and $l$ being dimensionless numbers and $a^{*}$ and $c^{*}$ being reciprocal lattice constants.

\subsection{\label{helicalB}Mathematica programs used in the present work} 
In order to determine the magnetic Bragg peaks in the reciprocal lattice plane, we use the following three Mathematica programs, ContourPlot for the equi-energy contour plot, FindMaximum for finding maximum, and ListVectorPlot3D for drawing the spin directions in the lattice points of the real space. 

\subsubsection{\label{helicalB1}Contour plot program}
This program is used to determine the overview on the positions of the magnetic Bragg peaks in the $(h,k,l)$ reciprocal lattice plane, where one of $h$, $k$, and $l$ are fixed. When the values of $J_{1}$, $J_{2}$, and $J_{3}$ are given, using the Mathematica program [ContourPlot], we can make a plot of the contours of $J(h,k,l)=a$ (constant) in the $(h,k,l)$ plane, where the parameter $a$ is changed appropriately such that the maximum value of $J(h,k,l)$ is obtained. This program is very useful to find the selection rule for the position of the magnetic Bragg peaks, as a function of $J_{1}$, $J_{2}$, and $J_{3}$. 

\subsubsection{\label{helicalB2}Finding maximum program} 
This program is used to determine the exact position of the magnetic Bragg peak in the $(h,k,l)$ reciprocal lattice plane, where one of $h$, $k$, and $l$ are fixed. For convenience, here we assume that $l$ is fixed such that $l$ = 0. Using the Mathematica program [FindMaximum], we find the maximum value of $J(h,k,l)$ for given $J_{2}$ and $J_{3}$, where $J_{1}$ (= -1) is fixed, and the regions of $h$ and $k$ are appropriately chosen. This program is very convenient when one wants to know how the position of the Bragg point changes as a function of $J_{2}$ and $J_{3}$. 

\subsubsection{\label{helicalB3}ListVectorPlot3D program} 
This program is used to draw the spin directions at each lattice sites in the real space. The spin structure depends on the values of {\bf Q}. 

\section{\label{sel}Selection rule for the magnetic Bragg peaks from simple analysis} 
\subsection{\label{selA}Distribution of magnetic Bragg peaks in the ($h$, $k$, $l$ = fixed integer) reciprocal lattice plane} 
The magnetic Bragg peaks can be located at the wavevector {\bf q}, where $J(h,k,l)$ takes a maximum. When $l$ = fixed integer, $J(h,k,l)$ is given by 
\begin{eqnarray*}
J(h,k,l)&=&8J_{1} (-1)^{l} \cos (\pi h)\cos (\pi k)  \\
&+&2J_{2} +2J_{3} \lbrack \cos (2\pi h)+\cos (2\pi k)\rbrack, 
\end{eqnarray*}
suggesting that the location of the magnetic Bragg peaks in the ($h$, $k$, $l$ = fixed integer) reciprocal lattice plane depends only on $J_{1}$ and $J_{3}$,and is independent of the $J_{2}$. The selection rule can be derived as follows. \\ 
(i) The distribution of the magnetic Bragg peaks is the same in the reciprocal lattice planes, ($h$, $k$, 1), ($h$, $k$, 3), ($h$, $k$, 5), ($h$, $k$, 7),... \\ 
(ii) The distribution of the magnetic Bragg peaks is the same in the reciprocal lattice planes , ($h$, $k$, 0), ($h$, $k$, 2), ($h$, $k$, 4), ($h$, $k$, 8),... 

\subsection{\label{selB}Magnetic Bragg peaks along the $l$ direction with $h$ = fixed integer and $k = 0$} 
When $k$ = 0 and $h$ is a fixed integer, $J$($h$ = fixed integer, $k$ = 0, $l$) is rewritten as 
\[
J(h,0,l)=8(-1)^{h} J_{1} \cos (\pi l)+2J_{2} \cos (2\pi l)+2J_{3} ,
\] 
which indicates that the location of the magnetic Bragg peaks along the ($h$ = fixed integer, $k$ = 0, $l$) direction depends only on $J_{1}$ and $J_{2}$, and is independent of $J_{3}$. The following selection rules can be derived.\\ 
(i) The location of the magnetic Bragg peaks along the $l$ direction is the same for (1, 0, $l$), (3, 0, $l$), (5, 0, $l$), (7, 0, $l$), $\cdots$. \\
(ii) The location of the magnetic Bragg peaks along the $l$ direction is the same for (0, 0, $l$), (2, 0, $l$), (4, 0, $l$), (8, 0, $l$), $\cdots$.  

\subsection{\label{selC}Magnetic Bragg peaks along the $l$ direction with $h=k=$ half inetgers} 
For $h=k=$ half-integer (= 1/2, 3/2, 5/2, $\cdots$), we have
\[ 
J(h,h,l)=2J_{2} \cos (2\pi l)-4J_{3} ,
\] 
leading to the appearance of the magnetic Bragg peaks appear at 
\[ 
l = 1/2, 3/2, 5/2, \cdots 
\]
for $h=k= 1/2, 3/2, 5/2, 7/2, \cdots$ .

\subsection{\label{selD}Magnetic Bragg peaks along the $(0,0,l)$ direction} 
For any $l$, $J(0,0,l)$ can be expressed by
\[ 
J(h=0,k=0,l)=J(\theta)=8J_{1}\cos (\theta )+2J_{2}\cos (2\theta)+4J_{3} .
\] 
with $\theta =\pi l$ . This indicates that the location of the magnetic Bragg peaks along the $(0,0,l)$ direction, is independent of $J_{3}$. The derivative of $J(0,0,l)$ with respect to $\theta$ is obtained as 
\[ 
\frac{dJ(\theta)}{d\theta}=-4\lbrack J_{1} +J_{2} \cos (\theta )\rbrack\sin (\theta )=0  .
\] 
Then we get the two solutions.\\
(i) Ferromagnetic or antiferromagnetic configurations
\[ 
\sin \theta =0,  \text{     }\theta = 0, \pi  . 
\]
where
\begin{eqnarray*}
J(\theta  = 0)  & = &8J_{1} +2J_{2} +4J_{3} \\
J(\theta  = \pi)& = &-8J_{1} +2J_{2} +4J_{3} . 
\end{eqnarray*}
(ii) Helical spin configuration
\[ 
\cos \theta =-J_{1} /J_{2}  , 
\theta=\theta_{0} = \arccos(-J_{1}/J_{2}), 
\] 
under the condition of $\left| J_{1}/J_{2}\right| <1$, where
\[ 
J(\theta =\theta_{0})=-4\frac{J_{1}^{2}}{J_{2}}-2J_{2}+4J_{3} . 
\]
The difference is calculated as
\begin{eqnarray*}
J(\theta  & = & \theta _{0} )-J(\theta =\pi )=-4J_{2} (\frac{J_{1} }{J_{2} } -1)^{2} ,  \\
J(\theta  & = & \theta _{0} )-J(\theta =0)=-4J_{2} (\frac{J_{1} }{J_{2} } +1)^{2} . 
\end{eqnarray*}
Then the helical spin order appears when
\begin{eqnarray*} 
J(\theta =\theta _{0} )>J(\theta =\pi ) , \\
J(\theta =\theta _{0} )>J(\theta =0 ) . 
\end{eqnarray*}
If the conditions $J_{2}<0$ and $\left| J_{1}/J_{2}\right|<1$ are satisfied, the magnetic Bragg peaks appear at the wavevectors denoted by 
\[
(h,k,l) = (0,0,1\pm\epsilon), (0,0,3\pm\epsilon), (0,0,5\pm\epsilon), (0,0,7\pm\epsilon),\cdots , 
\]
where $\epsilon$ is defined as
\[ 
\cos (\pi \epsilon )=\frac{J_{1} }{J_{2} } . 
\]

\subsection{\label{selE}Magnetic Bragg peaks along the $(h,0,0)$ direction} 
$J(h,k=0,l=0)$ and its derivative are given by 
\[ 
J(h,k=0,l=0)=8J_{1} \cos (\pi h)+2J_{3} \cos (2\pi h)+2J_{2} +2J_{3} ,
\]
and
\[ 
\frac{dJ(h,k=0,l=0)}{dh} =8\pi \lbrack J_{1} -J_{3} \cos (\pi h)]\sin (\pi h)  ,
\] 
respectively. Since $\left| J_{1}/J_{3}\right| >1$ in MnO$_{2}$, we have only $\sin(\pi h) = 0$. Since $J_{1}<0$, we have the following selection rule. The magnetic Bragg appears only at
\[
(1,0,0), (3,0,0), (5,0,0), (7,0,0), \cdots 
\] 
along the $h$-direction. 

\section{\label{det}Determination of $J_{2}$ and $J_{3}$ from experimental data on $\beta$-$\text{MnO}_{2}$}
\subsection{\label{detA}The ratio $J_{2}/J_{1}$} 

\begin{figure}
\includegraphics[width=8.0cm]{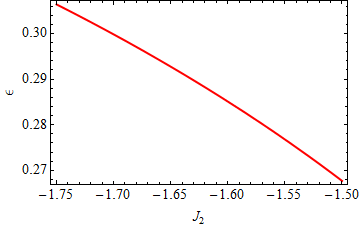}
\caption{\label{fig02}(Color online) Plot of $\epsilon$ as a function of $J_{2}$. $J_{1}=-1$}
\end{figure}

We now calculate the value of $\epsilon$, at which $J(h=1,k=0,l=2+\epsilon)$ takes a maximum, as a function of $J_{2}$, where $J_{1} = -1$. The value of $\epsilon$ is uniquely determined as a function $J_{2}$ as shown in Fig.~\ref{fig02}. When $J_{2} = -1.6039$, we find $\epsilon = 2/7$ (Yoshimori\cite{ref02}).

Experimentally, the magnetic Bragg peak is observed at $(h=1,k=0,l=2.29771)$ for $\beta$-MnO$_{2}$. Here we note that $l = 2.29771 = 2 + \epsilon$, where $\epsilon =2/7 + 0.0120$. The value of $l$ is slightly deviated from the value of 2 + 2/7, which is predicted by Yoshimori.\cite{ref02} $J(h=1,k=0,l)$ and its derivative are given by
\[ 
J(h=1,k=0,l)=-8J_{1} \cos (\pi l)+2J_{2} \cos (2\pi l)+4J_{3} ,
\]
and
\[ 
\frac{J(h=1,k=0,l)}{dl} =4\pi \lbrack 2J_{1} \sin (\pi l)-J_{2} \sin (2\pi l)\rbrack ,
\]
respectively. $J(h=1,k=0,l)$ has a maximum when
\[ 
\cos (\pi l)=\frac{J_{1} }{J_{2} } .
\]
or
\[ 
l= 0\pm\epsilon, 2\pm\epsilon, 4\pm\epsilon, 6\pm\epsilon,\cdots ,
\] 
where
\[ 
\cos (\pi \epsilon )=\frac{J_{1} }{J_{2} } .
\]
Since the magnetic Bragg peak appears at $l = 2.29771$, we have 
\[ 
\frac{J_{2} }{J_{1} } = 1.68469 ,
\] 
indicating that $J_{2}<0$ since $J_{1}<0$.

\subsection{\label{detB}Ratio $p_{3}=J_{3}/J_{1}$} 

\begin{figure}
\includegraphics[width=8.0cm]{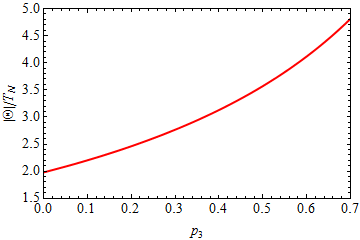}
\caption{\label{fig03}(Color online) Plot of $\left|\Theta\right|/T_{N}$ as a function of $p_{3}$, where $p_{2} = 1.68469$.}
\end{figure}

The ratio $J_{3}/J_{1}$ cannot be determined uniquely from the experimental results on the magnetic neutron scattering or x-ray magnetic scattering. The ratio $\left|\Theta\right|/T_{N}$ can be calculated as 
\begin{equation}
\frac{\left|\Theta\right|}{T_{N}} =\frac{J(0)}{J(\theta_{0})}=\frac{4+p_{2}+2p_{3}}{\frac{2}{p_{2}}+p_{2}-2p_{3} }  ,
\label{eq12}
\end{equation} 
where $p_{2}=J_{2}/J_{1}$, $p_{3}=J_{3}/J_{1}$, $\Theta$ is the Curie-Weiss temperature (the value is negative for $\beta$-MnO$_{2}$) and $T_{N}$ is the N\'{e}el temperature. The value of $p_{3}$ can be determined from the ratio $\left|\Theta\right|/T_{N}$ since $p_{2}$ is already determined. We make a plot of $\left|\Theta\right|/T_{N}$ as a function of $p_{3}$ in Fig.~\ref{fig03}, where $p_{2}=1.68469$.

As far as we know, there have been several reports on the experimental values of $\Theta$ and $T_{N}$ for $\beta$-MnO$_{2}$; $\Theta = -316$ K, $T_{N} = 84$ K [Bizette and B. Tsai (1949)],\cite{ref10} $\Theta = -1050$ K, $T_{N} = 92$ K [Ohama and Hamaguchi (1971)],\cite{ref03} $\Theta = -783$ K, $T_{N} = 92$ K [Sato et al.(2000)],\cite{ref08} respectively. Using the values of $T_{N}$, $\Theta$ and the value of $p_{2}$ (= 1.68469) determined above, the value of $p_{3}$ can be calculated as $p_{3}$ = 0.537, 0.986, 1.091 using the data of Bizette and Tsai,\cite{ref10} Sato et al.,\cite{ref08} and Ohama and Hamaguchi,\cite{ref03} respectively. 

We note that the boundary of the helical phase is described by (the derivation will be shown later) 
\begin{equation}
p_{3} = 1/p_{2} \text{ or } J_{2}J_{3}=J_{1}^{2}  .
\label{eq13}
\end{equation} 
The helical order can exist only when $p_{3}<1/p_{2}$. When $p_{2} = 1.68469$ for MnO$_{2}$, $p_{3}$ should be lower than 0.5935. In other words, the ratio $\left|\Theta\right|/T_{N}$ should be lower than 4.07853. Since $T_{N} = 92$ K, this means that $\left|\Theta\right|$ should be smaller than 375.2 K. The value of $\left|\Theta\right|$ by Bizette and Tsai seems to be reasonable, while the values of $\left|\Theta\right|$ obtained by Sato et al.\cite{ref08} and Ohama and Hamaguchi\cite{ref03} are much higher than 375.2 K. Note that Yoshimori\cite{ref02} predicts that the boundary is described by $p_{3}p_{2}=1/2$. This expression is not correct according to our calculation.

\section{\label{cal}Numerical calculation}
\subsection{\label{calA}The contour plot of $J(h,0,l)$ in the $(h,0,l)$ reciprocal lattice plane} 

\begin{figure}
\includegraphics[width=4.0cm]{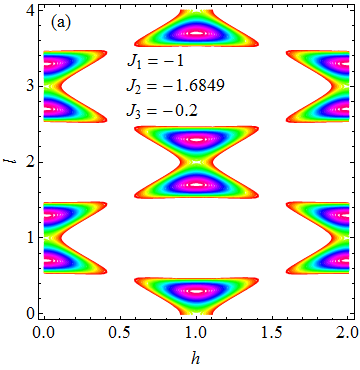}
\includegraphics[width=4.0cm]{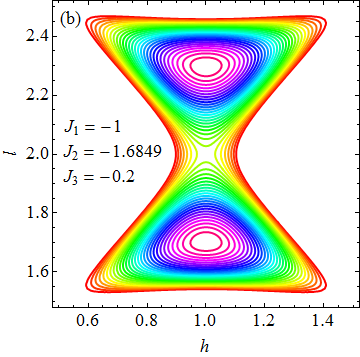}
\includegraphics[width=4.0cm]{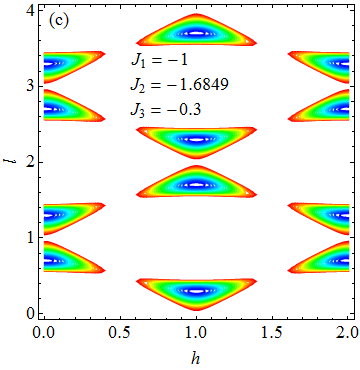}
\includegraphics[width=4.0cm]{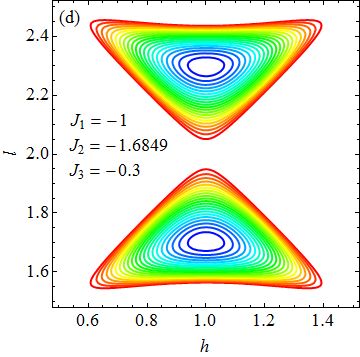}
\includegraphics[width=4.0cm]{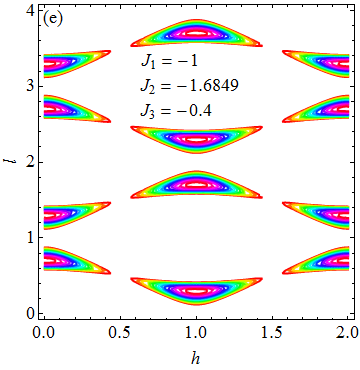}
\includegraphics[width=4.0cm]{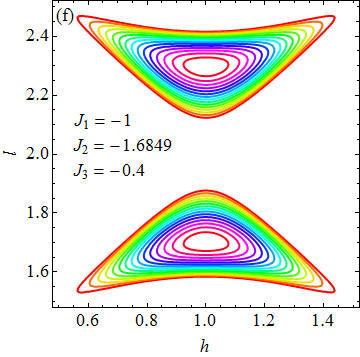}
\caption{\label{fig04}(Color online) (a)(b) Distribution of magnetic Bragg points in the $(h,k=0,l)$ reciprocal lattice plane. $J_{1} = -1$, $J_{2} = -1.68469$. $J_{3} = -0.2$. $p_{3}=J_{3}/J_{1} = 0.2$. (c)(d) Distribution of magnetic Bragg points in the $(h,k=0,l)$ reciprocal lattice plane. $J_{1} = -1$, $J_{2} = -1.68469$. $J_{3} = -0.3$. $p_{3}=J_{3}/J_{1} = 0.3$. (e)(f) Distribution of magnetic Bragg points in the $(h,k=0,l)$ reciprocal plane around $h=1$ and $l=2$. $J_{1}=-1$, $J_{2} = -1.68469$. $J_{3}=-0.4$. $p_{3}=J_{3}/J_{1}=0.4$.}
\end{figure}

Using the numerical calculation of the contour plot using the Mathematica, we find the location of the magnetic Bragg peaks in the $(h,0,l)$ reciprocal lattice, where 
\begin{eqnarray*}
J(h,0,l)&=&8J_{1} \cos (\pi h)\cos (\pi l) \\
&+&2J_{2} \cos (2\pi l)+2J_{3} \lbrack \cos (2\pi h)+1] , 
\end{eqnarray*}
takes a maximum, where $J_{1} = -1$, $J_{2}=p_{2}J_{1} = -1.68469$. Figures \ref{fig04}(a)-(f) show the contour plot of $J(h,0,l)$ in the reciprocal lattice plane of $(h,0,l)$, where $J_{3}=p_{3}J_{1}$ ($p_{3}<0.5935$) and $J_{3}$ is changed as a parameter ($J_{3}=-0.2, -0.3, -0.4$). As is predicted above, we find that the magnetic Bragg peaks appear at 
\begin{eqnarray*}
(0, 0, 1 \pm\epsilon), (0, 0, 3 \pm\epsilon), (0, 0, 5 \pm\epsilon), (0, 0, 7 \pm\epsilon),\dots ,  \\
(1, 0, 0 \pm\epsilon), (1, 0, 2 \pm\epsilon), (1, 0, 4 \pm\epsilon), (1, 0, 6 \pm\epsilon),\dots ,  \\
(2, 0, 1 \pm\epsilon), (2, 0, 3 \pm\epsilon), (2, 0, 5 \pm\epsilon), (2, 0, 7 \pm\epsilon),\dots ,  \\
\end{eqnarray*} 
where $\epsilon = 0.29771$ and is independent of $p_{3}$ ($p_{3}<0.5935$). Note that the essential results are independent of the choice of $p_{3}$ at least for $0.2\le p_{3}\le 0.4$.

\subsection{\label{calB}The contour plot of $J(h,k=h,l)$ in the $(h,k=h,l)$ reciprocal lattice plane} 
Using the contour plot of the $J(h,k=h,l)$, we find the location of the magnetic Braggs in the $(h,h,l)$ reciprocal lattice, where 
\[
J(h,h,l)=8J_{1} \cos ^{2} (\pi h)\cos (\pi l)+2J_{2} \cos (2\pi l)+4J_{3} \cos (2\pi h) .
\]
The location of the magnetic Bragg peaks is given by
\begin{widetext}
\begin{eqnarray*} 
(0, 0, 1\pm\epsilon), (0, 0, 3\pm\epsilon), (0, 0, 5\pm\epsilon), (0, 0, 7\pm\epsilon),\dots , \\
(1/2, 1/2, 1/2), (1/2, 1/2, 3/2), (1/2, 1/2, 5/2), (1/2, 1/2, 7/2),\dots , \nonumber \\
(1, 1, 1\pm\epsilon), (1, 1, 3\pm\epsilon), (1, 1, 5\pm\epsilon), (1, 1, 7\pm\epsilon),\dots , \\
(3/2, 3/2, 1/2) (3/2, 3/2, 3/2), (3/2, 3/2, 5/2), (3/2, 3/2, 7/2),\dots , 
\end{eqnarray*}
\end{widetext}
in the reciprocal lattice plane of $(h,k=h,l)$, where $\epsilon = 0.29771$.

\subsection{\label{calC}The contour plot in the $(h,k,l)$ plane with $l=2n$} 
Using the contour plot of the Mathematica, we find the location of the magnetic Braggs in the $(h,k,l=2n)$ reciprocal lattice, where 
\begin{eqnarray*}
J(h,k,l=2n)&=&8J_{1} \cos (\pi h)\cos (\pi k) \\
&+&2J_{3} \lbrack \cos (2\pi h)+\cos (2\pi k)\rbrack +2J_{2} ,
\end{eqnarray*}
which does not depend on the index $n$. Figures \ref{fig05}(a) show the location of magnetic Bragg peaks in the in-plane $(h,k,l= 2n)$ reciprocal lattice plane. We find that the magnetic Bragg peaks appear for $h+k =$ even. This selection rule is true for $l = 2n$ with any integer $n$.

\subsection{\label{calD}The contour plot in the $(h,k,l)$ plane with $l=2n+1$} 

\begin{figure}
\includegraphics[width=4.0cm]{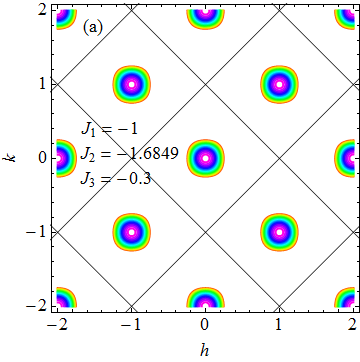}
\includegraphics[width=4.0cm]{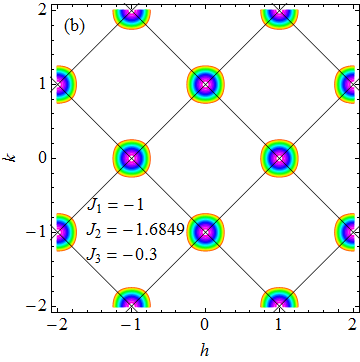}
\caption{\label{fig05}(Color online) (a) Magnetic Bragg points in the $(h,k,l=2n)$ reciprocal lattice plane. (b) Magnetic Bragg points in the $(h,k,l=2n+1)$ reciprocal lattice plane. $J_{1} = -1$, $J_{2} = -1.68469$. $J_{3} = -0.3$. $p_{3} = 0.3$.}
\end{figure}

Using the contour plot of the Mathematica, we find the location of the magnetic Braggs in the $(h,k,l=2n+1)$ reciprocal lattice, where 
\begin{eqnarray*}
J(h,k,l&=&2n+1)=-8J_{1} \cos (\pi h)\cos (\pi k)) \\
&+&2J_{3} \lbrack \cos (2\pi h)+\cos (2\pi k)\rbrack +2J_{2} , 
\end{eqnarray*}
which does not depend on the index $n$. Figure \ref{fig05}(b) show the location of the magnetic Bragg peaks in the in-plane $(h,k, l=2n+1)$ reciprocal lattice plane, where $n$ is the integer. The magnetic Bragg peaks appear for $h+k=$ odd for $l=2n+1$.

\section{\label{phase}Phase diagram in the $(J_{2},J_{3})$ plane} 
The phase diagram for $J_{1} = -1$ can be determined using the programs of the ContourPlot and FindMaximum. The phase diagram consists of the four phases; (i) the helical order along the $c$ axis, (ii) the helical order along the $a$ axis, (iii) the ordered phase with $h = 1/2$, $k = 0$, and $l = 1/2$, and (iv) the ordered phase with $h = 0$, $k = 0$, $l = 1$ is shown in the phase diagram Fig.~\ref{fig06}. 

\subsection{\label{phaseA}Ordered phases}
\subsubsection{Helical order along the $c$ axis} 
The spin vectors in the same $ab$ plane are parallel, i.e., $h = 0$ and $k = 0$. They screw along the $c$ axis. We find the maximum of 
\begin{eqnarray*}
J(h=0,k=0,l)&=&4J_{3} +8J_{1} \cos (l\pi )  \\
&+&2J_{2} \cos (2l\pi ) ,  
\end{eqnarray*}
by taking the derivative of $J(h=0,k=0,l)$ with respect to $l$. The condition of the local maximum is given by 
\[
\cos (\pi l)=-\frac{J_{1}}{J_{2}} ,
\]
where $\left| J_{1}/J_{2}\right| <1$. Then the maximum of $J(h=0,k=0,l)$ is given by
\begin{equation}
J_{\max } (h=0,k=0,l)=4J_{3} -4\frac{J_{1} ^{2} }{J_{2} } -2J_{2}  ,
\label{EQN15}
\end{equation}
for the helical order with ($h=0$, $k=0$, and $l$) where $l=0.70228$,

\subsubsection{Helical order along the $a$ axis} 
The helical spin order where the spin vector in the same \textit{bc} plane are parallel and they screw along the $a$ axis. We find the maximum of 
\[
J(h,k=0,l=1)=2J_{2} +2J_{3} +8J_{1} \cos (\pi h)+2J_{3} \cos (2\pi h) ,
\]
by taking the derivative of $J(h,k=0,l=0)$ with respect to $h$. The condition of the local maximum is given by 
\[
\cos (\pi h)=-\frac{J_{1} }{J_{3} } ,
\] 
where $\left| J_{1}/J_{3}\right| <1$. The maximum value is
\begin{equation}
J_{\max } (h,k=0,l=1)=2J_{2} -4\frac{J_{1} ^{2} }{J_{3} } .
\label{EQN16}
\end{equation}

\subsubsection{The ordered phase with $h = 0$, $k = 0$, and $l = 1$} 
$J(h=0,k=0,l=1)$ is given by
\begin{equation}
J(h=0,k=0,l=1)=-8J_{1} +2J_{2} +4J_{3} .
\label{EQN17}
\end{equation}

\subsubsection{The ordered phase with $h=1/2$, $k=0$, and $l=1/2$} 
This phase corresponds to the MnF$_{2}$-type antiferromagnetic structure (named by Yoshimori\cite{ref02}). $J(h=1/2,k=0,l= 1/2)$ is evaluated as 
\begin{equation}
J(h=1/2,k=0,l=1/2)=-2J_{2} .
\label{EQN18}
\end{equation}

\subsection{\label{phaseB}Boundaries between ordered phases}
\subsubsection{The phase boundary between the spin order with the phase ($h=1/2$, $k=0$, and $l=1/2$) and the helical spin order along the $c$ axis}
The difference of $J(h,k,l)$ between the helical spin order along the $c$ axis and the phase $(h=1/2,k=0,l=1/2)$ is given by
\begin{eqnarray*}
J_{\max }(h=0,k=0,l)-J(h=1/2,k=0,l=1/2)  \\
=4J_{2} -4\frac{J_{1} ^{2}}{J_{3}} . 
\end{eqnarray*}
So the helical ordered phase is energetically favorable when
\[ 
J_{2} >\frac{J_{1} ^{2} }{J_{3} } ,
\] 
with the condition of $\left| J_{1}/J_{2}\right| <1$. When $J_{3}<0$, this inequality can be rewritten as 
\begin{equation}
J_{2} J_{3} <J_{1} ^{2} ,
\label{EQN19}
\end{equation}
or
\[ 
p_{2}p_{3} <1 .
\] 

\subsubsection{The phase boundary between the spin order with ($h=0$, $k=0$, $l=1$) and the spin order with ($h=1/2$, $k= 0$, $l=1/2$)} 
The difference of $J(h,k,l)$ between the $(h=0,k=0,l=1)$ and the phase $(h=1/2,k=0,l=1/2)$ is given by 
\begin{eqnarray*}
J(h=1/2,k=0,l=1/2)-J(h=0,k=0,l=1)  \\
=-4J_{2} -4J_{3} +8J_{1} .
\end{eqnarray*}
When 
\begin{equation}
2J_{1} >J_{2} +J_{3} , 
\label{EQN20}
\end{equation}
the phase $(h=1/2,k=0,l= 1/2)$ is energetically favorable. When 
\begin{equation}
2J_{1} <J_{2} +J_{3} .
\label{EQN21}
\end{equation}
the phase $h = 0$, $k = 0$, $l = 1$) is energetically favorable. 

\subsubsection{The phase boundary between the helical order along the $a$ axis and the phase ($h=k=1/2$, and $l=1$)} 
The difference of $J(h,k,l)$ between the $(h=0,k=0,l=1)$ and the helical order along the $a$ axis is given by 
\begin{eqnarray*}
J_{\max } (h,k=0,l=1)-J(h=0,k=0,l=1)  \\
=-4J_{3} (\frac{J_{1} }{J_{3} } -1)^{2} .
\end{eqnarray*}
Then the helical order is stable for $J_{3}<-1$. This inequality is satisfied for the condition of the helical order 
\[
\left| J_{1}/J_{3}\right| <1.
\] 

\subsection{\label{phaseC}Multicritical point ($J_{2}=-1$ and $J_{3}=-1$) and phase transition lines} 

\begin{figure}
\includegraphics[width=8.0cm]{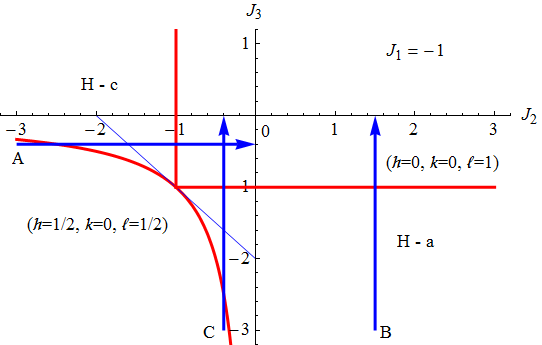}
\caption{\label{fig06}(Color online) Scans A, B, and C in the phase diagram ($J_{2}$, $J_{3}$) to determine the location of the magnetic Bragg peaks in the $(h,k,l)$ reciprocal space. $J_{1} = -1$. Note that $J_{2}=-1.68469$ and $J_{3}=-0.537$ for $\beta$-MnO$_{2}$ (denoted by an open circle).}
\end{figure}

All four ordered phases merge only at the multicritical point at ($J_{2}=-1$ and $J_{3}=-1$), where the line given by $2J_{1} =J_{2} +J_{3}$ and the curve $J_{2} J_{3} =J_{1} ^{2}$ intersect with other (see the phase diagram of $J_{3}$ vs $J_{2}$). Note that $J_{1} = -1$. As will be shown later, the line ($2J_{1}=J_{2}+J_{3}$) is of the first-order. The lines denoted by$J_{2}=-1$ and $J_{3}=-1$ are of the second-order. \\
(1) Using the program of finding maximum in $J({\bf q})$ for each point ($J_{2}$, $J_{3}$) in the phase digram, we find the local maximum point $(h,0,l)$ in the $h$-$l$ plane. The location of these points is plotted as a function of $J_{2}$ when $J_{3}$ is fixed and as a function of $J_{3}$ when $J_{2}$ is fixed.\\
(2) Using the program of finding maximum in $J({\bf q})$ for each point ($J_{2}$, $J_{3}$) in the phase diagram,  we find the local maximum point $(h,k,0)$ in the $h$-$k$ plane. These points are plotted as a function of $J_{2}$ when $J_{3}$ is fixed and as a function of $J_{3}$ when $J_{2}$ is fixed. 

\subsubsection{Scan A} 

\begin{figure}
\includegraphics[width=8.0cm]{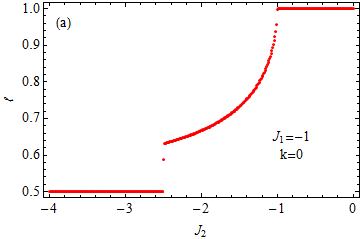}
\includegraphics[width=8.0cm]{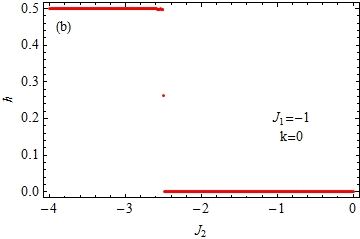}
\caption{\label{fig07}(Color online) (a) $l$ vs $J_{2}$ and (b) $h$ vs $J_{2}$ along the scan A ($-4\le J_{2}\le 0$ and $J_{3}=-0.4$). $k=0$. $J_{1}=-1$.}
\end{figure}

In the Scan A (see Fig.~\ref{fig06}), we choose $J_{3} = -0.4$ and $J_{1}=-1$. $J_{2}$ is changed as a parameter between $-4.0$ and 0. We make a plot of $l$ vs $J_{2}$ in Fig.~\ref{fig07}(a). The discontinuity in $l$ vs $J_{2}$ occurs at $J_{2} = -2.5$ [the phase boundary (first order) between the phase with $(h=1/2,k=0,l=1/2)$ and the helical order along the $c$ axis], where the relation $J_{2}J_{3}=J_{1}^{2}$ is satisfied ($J_{1}=-1$). We find that the relation of $l$ vs $J_{2}$ is well described by 
\begin{equation}
\cos (\pi l)=-\frac{J_{1} }{J_{2} } ,
\label{EQN22}
\end{equation}
for $-2.5\le J_{2}\le -1$ (the helical phase along the $c$ axis). For example, when $l=0.8$, we have $J_{2}=-1.24$ as shown in Fig.~\ref{fig07}(a). We note that $l$ is equal to 1 at $J_{2} = -1$ [the phase boundary(second order) between the helical phase along the $c$ axis and the phase with $(h=0,k=0,l=1)$] and remain unchanged ($l=1$) for $-1\le J_{2}\le 0$. We also make a plot of $h$ vs $J_{2}$ in Fig.~\ref{fig07}(b). The value of $h$ undergoes a sudden change from $h=0.5$ to 0 at $J_{2} = -2.5$ (at the phase boundary between the phase with $(h=1/2,k=0,l=1/2)$ and the helical order along the $c$ axis) and remains constant ($h=0$) for $-2.5\le J_{2}\le 0$. Note that there is no change of $h$ at $J_{2}=-1$. 

\subsubsection{Scan B} 

\begin{figure}
\includegraphics[width=8.0cm]{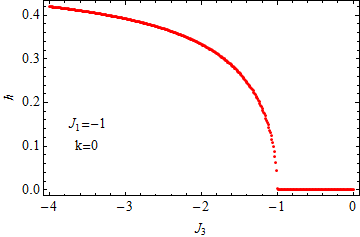}
\caption{\label{fig08}(Color online) $h$ vs $J_{3}$ along the scan B ($-4\le J_{3}\le 0$ and $J_{2}=1.5$). $J_{1}=-1$. $k = 0$ and $l = 1$.}
\end{figure}

In the Scan B (see Fig.~\ref{fig06}), we choose $J_{2} = 1.5$ and $J_{1} = -1$. $J_{3}$ is changed as a parameter between -4.0 and 0. We make a plot of $h$ vs $J_{2}$ in Fig.~\ref{fig08}. The value of $h$ decreases with increasing $J_{3}$ and reduces to zero at $J_{3} = -1$ [the phase boundary (second-order) between the helical order along the $a$ axis and the phase with $(h=0,k=0,l=1)$]. We find that the relation of $h$ vs $J_{3}$ is well described by 
\begin{equation}
\cos (\pi h)=-\frac{J_{1} }{J_{3} } .
\label{EQN23}
\end{equation}
The value of $h$ is independent of $J_{2}$. For example, when $h = 0.2$, we have $J_{3} = -1.236$. The plot of $l$ vs $J_{2}$ (which is not shown here) indicates that $l$ remains constant ($l = 1$) for $-4\le J_{3}\le 0$, where $J_{2} = 1.5$. 

\subsubsection{Scan C} 

\begin{figure}
\includegraphics[width=8.0cm]{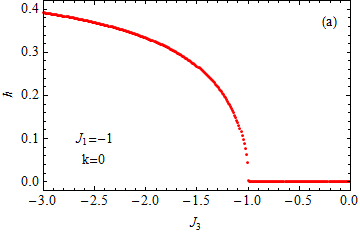}
\includegraphics[width=8.0cm]{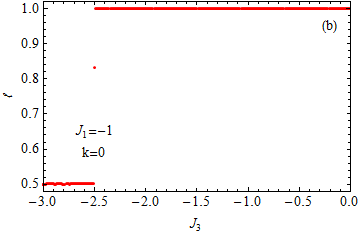}
\caption{\label{fig09}(Color online) (a) $h$ vs $J_{3}$ and (b) $l$ vs $J_{3}$ along the scan C ($-3.0\le J_{3}\le 0$ and $J_{2}=-0.4$). $J_{1}=-1$. $k=0$.}
\end{figure}

In the Scan C, we choose $J_{2} = -0.4$ and $J_{1} = -1$. $J_{3}$ is changed as a parameter between -3.5 and 0. We make a plot of $h$ vs $J_{3}$ and $l$ vs $J_{3}$ in Figs.~\ref{fig09}(a) and (b). As shown in Fig.~\ref{fig09}(a), the value of $h$ decreases with increasing $J_{3}$, showing an abrupt decrease at $J_{3} = -2.5$ (which is denoted by $J_{2}J_{3}=J_{1}^{2}$) [the phase boundary (first-order) between the phase with $(h=1/2,k=0,l=1/2)$ and the helical order along the $a$ axis]. The value of $h$ decreases with further increasing $J_{3}$, following Eq.(\ref{EQN23}), and reduces to zero at $J_{3} = -1$ [the phase boundary (second-order) between the helical order along the $a$ axis and the phase with $(h=0,k=0,l=1)$]. 

In Fig.~\ref{fig09}(b), the value of $l$ increases with increasing $J_{3}$. It shows an abrupt change from $l = 1/2$ to $l = 1$ at $J_{3} = -2.5$ at the phase boundary (first-order) between the phase with $(h=1/2,k=0,l=1/2)$ and the helical state along the $a$ axis. 

\section{\label{con}Contour plot of $(h,0,l)$ for typical points in the $(J_{2},J_{3})$ phase diagram} 

\begin{figure}
\includegraphics[width=8.0cm]{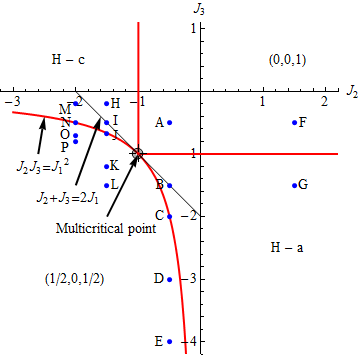}
\caption{\label{fig10}(Color online) Phase diagram of $J_{3}$ vs $J_{2}$: $J_{1}=-1$.$H$-$a$: helical order along the $a$ axis.$H$-$c$: helical order along the $c$ axis. The ordered phase with $(h=1/2,k=0,l=1/2)$. The ordered phase with $(h=0,k=0,l= 1)$. The multicritical point is at ($J_{2} = -1$, $J_{3} = -1$). The calculations of the contour plots are made at the points (A, B, C, $\cdots$, P). The line $J_{2}J_{3}=J_{1}^{2}$ is the first-order phase boundary. The line ($J_{2}=-1$, $J_{3}>-1$) and the line ($J_{3}=-1$, $J_{2}>-1$) are the second-order phase boundary.}
\end{figure}

Here we show the contour plot of $J(h,0,l)$ for the typical points in the ($J_{2}$, $J_{3}$) phase diagram (see Fig.~\ref{fig10}). The maximum points in the countour plot of $J(h,0,l)$) correspond to the magnetic Bragg peaks in the $(h,0,l)$ plane. 

\subsection{\label{conA}The points A, B, C, D, and E with $J_{1}=-1.0$ and $J_{2}=-0.5$} 

\begin{figure}
\includegraphics[width=4.0cm]{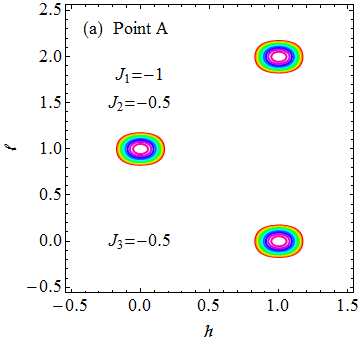}
\includegraphics[width=4.0cm]{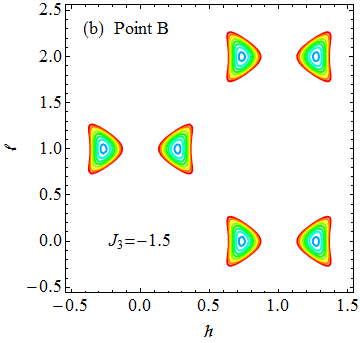}
\includegraphics[width=4.0cm]{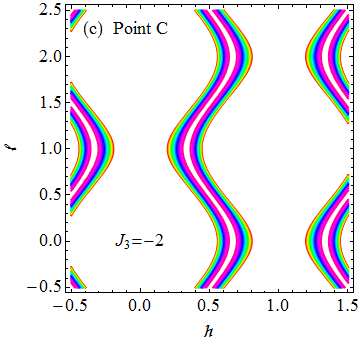}
\includegraphics[width=4.0cm]{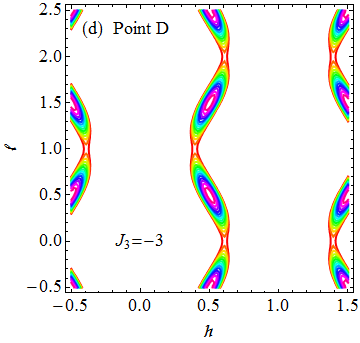}
\includegraphics[width=4.0cm]{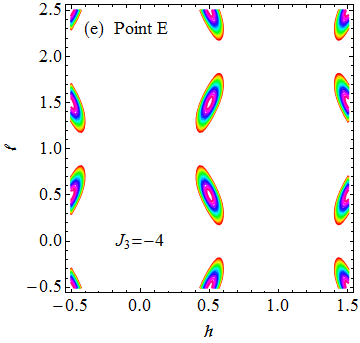}
\caption{\label{fig11}(Color online) Distribution of Bragg peaks in the $(h,0,l)$ reciprocal lattice plane. $J_{1} = -1$. $J_{2} = -0.5$, and $J_{3}$ is changed as a parameter. (a) Point A ($J_{3}=-0.5$), (b) Point B ($J_{3}=-1.5$), (c) Point C ($J_{3}=-2.0$), (d) Point D ($J_{3} = -3.0$), and (e) Point E ($J_{3} = -4.0$).}
\end{figure}

Typical contour plot of $(h,0,l)$ at the points A, B, C, D, and E in the ($J_{2}$, $J_{3}$) phase diagram are shown in Fig.~\ref{fig11}. The point A ($J_{3}=-0.5$) is in the phase with ($h$ = 0, $k$ = 0, $l$ = 1). The point B ($J_{3}$ = -1.5) is in the helical phase with the $a$ axis. The point C ($J_{3} = -2$) is the phase boundary between the phase with $(h=0,k=0,l=1)$ and the helical phase with the $a$ axis. The points D ($J_{3} = -3.0$) and E ($J_{3} = -4.0$) are in the phase with $(h=1/2,k=0,l=1/2)$. 

\subsection{\label{conB}The points F and G with $J_{1}=-1.0$ and $J_{2}=1.5$.} 

\begin{figure}
\includegraphics[width=4.0cm]{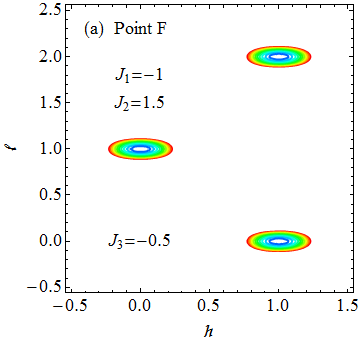}
\includegraphics[width=4.0cm]{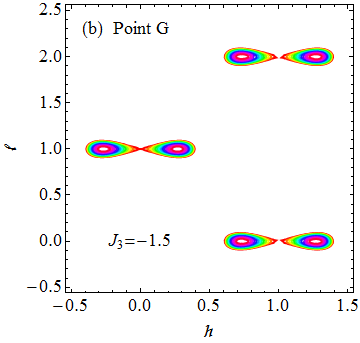}
\caption{\label{fig12}(Color online) Distribution of Bragg peaks in the $(h,0,l)$ reciprocal lattice plane. $J_{1} = -1$. $J_{2} = 1.5$, and $J_{3}$ is changed as a parameter. (a) F ($J_{3}=-0.5$) and (b) G ($J_{3}=-1.5$). }
\end{figure}

Typical contour plot of $(h,0,l)$ at the points F and G in the ($J_{2}$, $J_{3}$) phase diagram are shown in Fig.~\ref{fig12}. The point F ($J_{3} = -0.5$) is in the phase with $(h=0,k=0,l=1)$. The point G ($J_{3}=-1.5$) is in the helical phase with the $a$ axis. 

\subsection{\label{conC}The points H, I, J, K, and L with $J_{1}=-1.0$ and $J_{2}=-1.5$} 

\begin{figure}
\includegraphics[width=4.0cm]{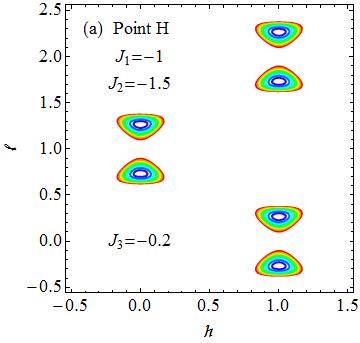}
\includegraphics[width=4.0cm]{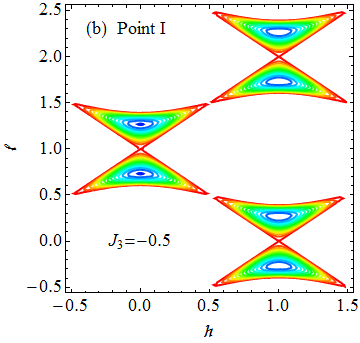}
\includegraphics[width=4.0cm]{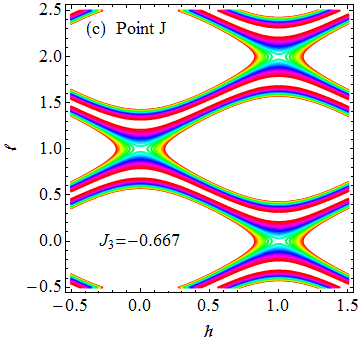}
\includegraphics[width=4.0cm]{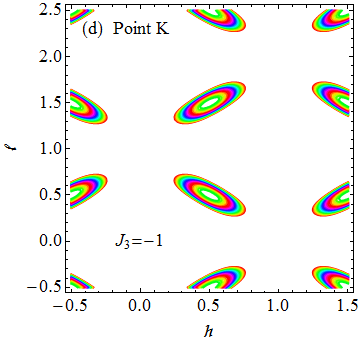}
\includegraphics[width=4.0cm]{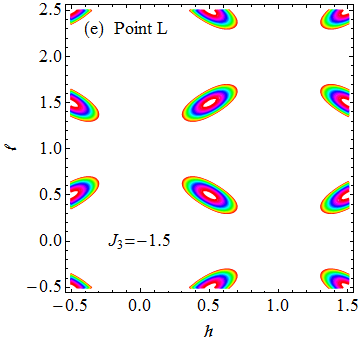}
\caption{\label{fig13}(Color online) Distribution of Bragg peaks in the $(h,0,l)$ reciprocal lattice plane. $J_{1}=-1$. $J_{2}=-1.5$, and $J_{3}$ is changed as a parameter. (a) H ($J_{3}=-0.2$), (b) I ($J_{3}=-0.5$), (c) J ($J_{3}=-0.667$), (d) K ($J_{3}=-1.2$), and (e) L ($J_{3}=-1.5$).}
\end{figure}

Typical contour plot of $(h,0,l)$ at the points H, I, J, K, and K in the ($J_{2}$, $J_{3}$) phase diagram are shown in Fig.~\ref{fig13}. The point H ($J_{3}=-0.2$) is in the helical phase along the $c$ axis. The point I ($J_{3}=-0.5$) is in the helical phase with the $c$ axis and is on the line ($J_{2}+J_{3}=2J_{1}$). The point J ($J_{3}=-0.667$) is on the phase boundary between the helical phase with the $c$ axis and the phase with $(h=1.2,k=0,l=1/2)$. The points K ($J_{3}=-1.2$) and L ($J_{3}=-1.5$) are in the phase with $(h=1/2,k=0,l= 1/2)$. 

\subsection{\label{conD}The points M, N, O, and P with $J_{1}=-1.0$ and $J_{2}=-2$} 

\begin{figure}
\includegraphics[width=4.0cm]{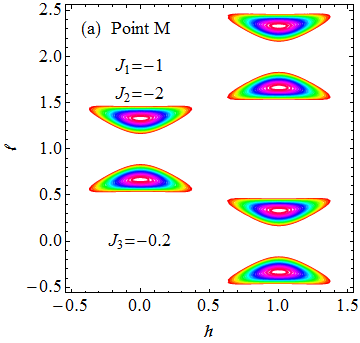}
\includegraphics[width=4.0cm]{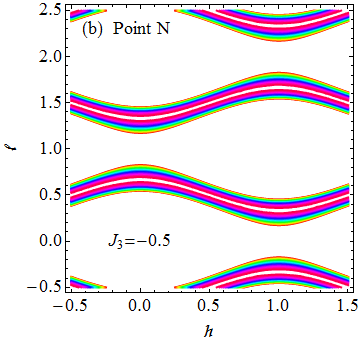}
\includegraphics[width=4.0cm]{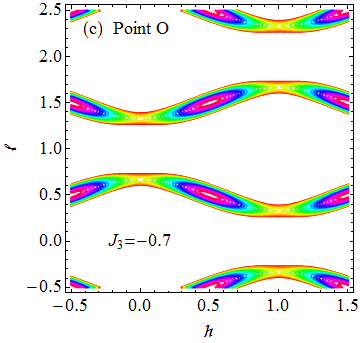}
\includegraphics[width=4.0cm]{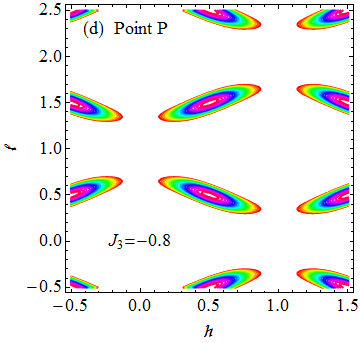}
\caption{\label{fig14}(Color online) Distribution of Bragg peaks in the $(h,0,l)$ reciprocal lattice plane. $J_{1} = -1$. $J_{2} = -2$, and $J_{3}$ is changed as a parameter. (a) M ($J_{3} = -0.2$), (b) N ($J_{3} = -0.5$), (c) O ($J_{3} = -0.7$), and (d) P ($J_{3} = -0.8$).}
\end{figure}

Typical contour plot of $(h,0,l)$ at the points M, N, O, and P in the ($J_{2}$, $J_{3}$) phase diagram are shown in Fig.~\ref{fig14} The point M ($J_{3} = -0.2$) is in the helical phase along the $c$ axis. The point N ($J_{3} = -0.5$) is on the phase boundary ($J_{2}+J_{3}=2J_{1}$) between the helical phase with the $c$ axis and the phase with $(h=1.2,k=0,l=1/2)$. The points O ($J_{3}=-0.7$) and P ($J_{3}=-0.8$) are in the phase with $(h=1/2,k=0,l=1/2)$. 

\section{\label{3D}3D spin structures} 
What is the three dimensional (3D) spin structure which is characterized with the wavevector of the magnetic Bragg peaks? The vector of spin at the site ${\bf R}_{i}$ of the real lattice space is given by 
\[
{\bf S}_{i}=S\lbrack\cos({\bf Q}\cdot {\bf R}_{i}){\bf e}_{x}+\sin({\bf Q}\cdot {\bf R}_{i}\phi ){\bf e}_{y} \rbrack
\] 
where $S = 3/2$, we assume that the phase factor $\phi$ is equal to zero, and {\bf Q} is defined as 
\[
{\bf Q}=(ha^{*},ka^{*},lc^{*}) .
\] 

\subsection{\label{3DA}Phase with $(h=0,k=0,l=1)$} 

\begin{figure}
\includegraphics[width=4.0cm]{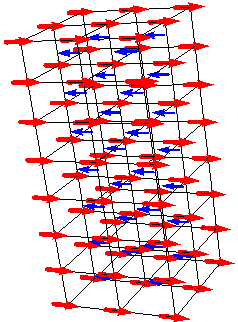}
\caption{\label{fig15}(Color online) structure in the phase $(h=0,k=0,l=1)$.}
\end{figure}

A typical spin structure for the phase with (0, 0, 1) is described in Fig.~\ref{fig15}.

\subsection{\label{3DB}Phase with $(h=1/2,k=0,l=1/2)$} 

\begin{figure}
\includegraphics[width=4.0cm]{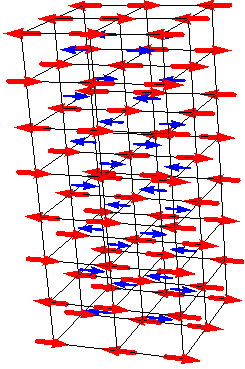}
\caption{\label{fig16}(Color online) Spin structure in the phase $(h=1/2,k=0,l=1/2)$.}
\end{figure}

A typical spin structure for the phase with (1/2, 0, 1/2) is described in Fig.~\ref{fig16}.

\subsection{\label{3DC}Helical ordered phase along the $c$ axis; $(0,0,l)$} 

\begin{figure}
\includegraphics[width=4.0cm]{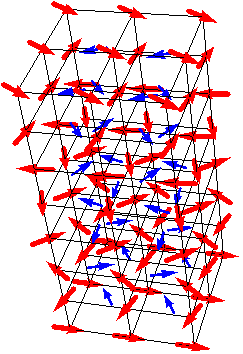}
\caption{\label{fig17}(Color online) Incommensurate spin structure for the helical order along the $c$ axis with the phase (0, 0, 0.7023). }
\end{figure}

The value $l$ is obtained as $l=0.7023$ for $J_{2} =-1.68469$ from Eq. (\ref{EQN22}). A typical spin structure is incommensurate with the lattice structure and described by Fig.~\ref{fig17}.

\subsection{\label{3DD}Helical ordered phase along the $c$ axis (Yoshimori); commensurate structure}

\begin{figure}
\includegraphics[width=4.0cm]{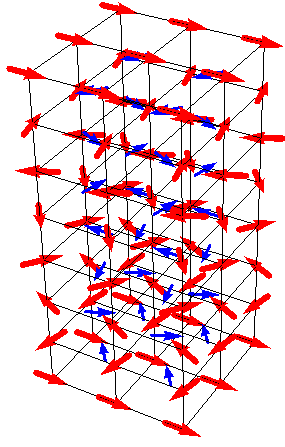}
\caption{\label{fig18}(Color online) Commensurate spin structure for the helical order along the c axis. with the phase $(0, 0, l=5/7)$, which is proposed by Yoshimori.\cite{ref02}}
\end{figure}
 
A typical spin structure with $l=0.71421$ is commensurate with the lattice structure and described by Fig.~\ref{fig18}. The magnetic unit cell along the $c$ axis is seven times larger than the unit cell of the crystal unit cell. 

\subsection{\label{3DE}Helical ordered phase along the $a$ axis: $(h,0,1)$} 

\begin{figure}
\includegraphics[width=8.0cm]{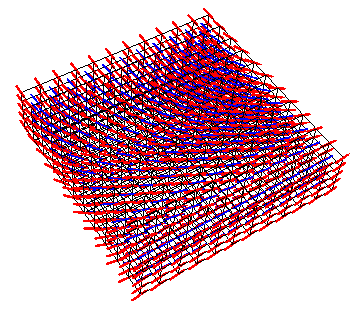}
\caption{\label{fig19}(Color online) Spin structure for the helical order along the $a$ axis with ($h_{max}$ = 2/3, 0, 1). $J_{1}=-1$, and $J_{3}=-2.0$.}
\end{figure}

When $J_{3}=-2.0$ and $J_{1}=-1$, $h$ is equal to 2/3 from Eq. {\ref{EQN23}). Then a typical spin structure is described by Fig.~\ref{fig19}.

\section{\label{dis}Discussion: nature of exchange interactions in the rutile-type $\beta$-$\text{MnO}_{2}$} 
\subsection{Overview}

\begin{figure}
\includegraphics[width=8.0cm]{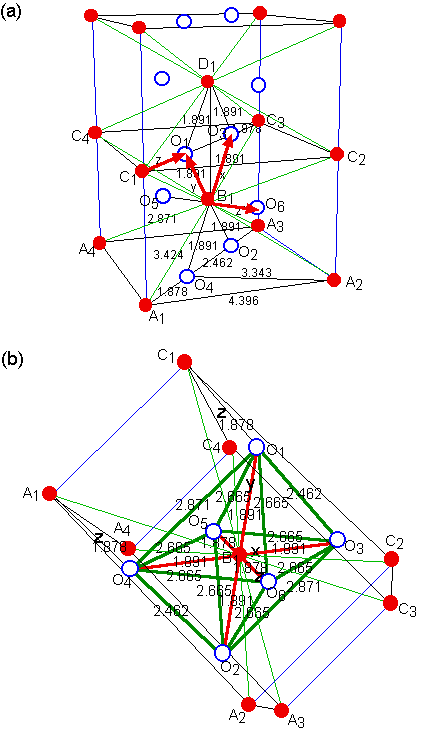}
\caption{\label{fig20}(Color online) (a) Crystal structure of $\beta$-MnO$_{2}$. $\angle$A$_{1}$O$_{4}$B$_{1}=\angle$C$_{1}$O$_{1}$B$_{1} = 130.62^{\circ}$. $\angle$A$_{1}$O$_{4}$A$_{2} = 111.60^{\circ}$. $\angle$B$_{1}$O$_{3}$D$_{1} = 98.77^{\circ}$. The distance (in the units of $\AA$) is denoted by numerical value. (b) Distorted octahedra of $\beta$-MnO$_{2}$. $\angle$O$_{1}$B$_{1}$O$_{4} = 98.77^{\circ}$, $\angle$O$_{1}$B$_{1}$O$_{3} = 81.225^{\circ}$, $\angle$O$_{5}$O$_{1}$O$_{6} = 89.590^{\circ}$, $\angle$O$_{1}$O$_{6}$O$_{5} = 45.205^{\circ}$, and $\angle$O$_{1}$B$_{1}$O$_{5} = \angle$O$_{4}$B$_{1}$O$_{5} = 90.0^{\circ}$. The distance (in the units of $\AA$) is denoted by numerical value.}
\end{figure}
 
We show that the magnetic phase diagram with $J_{1} = -1$ is uniquely determined by the combination of $J_{2}$ and $J_{3}$, where the sign of $J_{3}$ is mainly negative and the sign of $J_{2}$ is changed between negative to positive. For $\beta$-MnO$_{2}$, we have $p_{2} = 1.68469$ and as $p_{3} = 0.537$, leading to $J_{2} = -1.68469$ and $J_{3} = -0.537$ when $J_{1} = -1$. This point ($J_{2}$, $J_{3}$) for $\beta$-MnO$_{2}$ is in the helical order along the $c$ axis, as shown in the magnetic phase diagram (Fig.~\ref{fig06}). 

In Fig.~\ref{fig20}(a), we show the structure of $\beta$-MnO$_{2}$. Figure \ref{fig20}(b) shows the distorted octahedron where one Mn$^{4+}$ ion (cation) at the point B$_{1}$, is surrounded by six O$^{2-}$ ions (anion) at the points O$_{1}$, O$_{2}$, O$_{3}$, O$_{4}$, O$_{5}$, and O$_{6}$. The cation-cation separation (A$_{1}$C$_{1}$, B$_{1}$D$_{1}$) along a [001] axis is considerably smaller ($2.871 \AA$). Because of the distorted octahedron formed by O$^{2-}$ ions in the vicinity of Mn$^{4+}$ ion, the ground orbital state of Mn$^{4+}$ ion ($3d^{3}$, $L=3$ and $S=3/2$) is split into the $t_{2g}$ ($d\epsilon$) level [lower energy level, triple degenerate; $d(xy)$, $d(yz)$, and $d(zx)$ states] and the $e_{g}$ ($d\gamma$) level [upper energy, double degenerate; $d(3z^{2}-r^{2})$, $d(x^{2}-y^{2})$)]. The $t_{2g}$ state is occupied by three electrons with spin up state $\mid +\rangle$. The $\epsilon_{g}$ state is empty. As a result, the ground state is orbital singlet, indicating that the orbital angular momentum is quenched. In $\beta$-MnO$_{2}$, the $e_{g}$ electrons are responsible for the metallic conduction, while the localized $t_{2g}$ electrons are responsible for the magnetism. In Fig.~\ref{fig20}(b), the $(3z^{2}-r^{2})$ orbital axis at the point B$_{1}$ coincides with O$_{5}$B$_{1}$O$_{6}$ (the $z$ axis). The $d(x^{2}-y^{2})$ axes coincide with O$_{3}$B$_{1}$O$_{4}$ (the $x$ axis) and the O$_{1}$B$_{1}$O$_{2}$ (the $y$ axis). The $x$ and $y$ axes are perpendicular to the $z$ axis. However, the $x$ axis is not perpendicular to the $y$ axis ($98.78^{\circ}$, $81.22^{\circ}$). 

\subsection{\label{disA}Origin of the direct exchange interaction $J_{2}$} 
We consider the exchange interaction $J_{2}$ between the points B$_{1}$ and D$_{1}$ (or between the points A$_{1}$ and C$_{1}$). The distance B$_{1}$D$_{1}$ (A$_{1}$C$_{1}$) is $2.871 \AA$. In Fig.~\ref{fig20}(b), one $d(xy)$ orbital ($t_{2g}$) from the point B$_{1}$ bisects the angle  $\angle$O$_{1}$B$_{1}$O$_{3}$ and meets at the middle point of the edge O$_{1}$O$_{3}$. The other $d(xy)$ orbital from the point D$_{1}$ (the center of neighboring octahedron) bisects the angle  $\angle$O$_{1}$D$_{1}$O$_{3}$ and meets at the middle point of the edge O$_{1}$O$_{3}$. According to Goodenough,\cite{ref11} the predominant interactions between neighboring cations whose cation-ccupied octahedra share an edge, are assumed to be direct cation-cation exchange interaction. Thus the interaction $J_{2}$ is antiferromagnetic (Heitler-London type) for $\beta$-MnO$_{2}$. 

\subsection{\label{disB}Origin of superexchange interaction $J_{1}$} 
We consider the interaction between Mn$^{4+}$ at the point B$_{1}$ and the Mn$^{4+}$ at the point C$_{1}$ as shown in Figs.~\ref{fig20} (a) and (b), where the distance C$_{1}$B$_{1}$ is $3.424 \AA$. According to Goodenough,\cite{ref11} this interaction ($J_{1}$) is a superexchange one since the cation-occupied octahedra share a common corner (point O$_{1}$). The angle C$_{1}$O$_{1}$B$_{1}$ is equal to $130.62^{\circ}$, this interaction is antiferromagnetic for $\beta$-MnO$_{2}$. According to Goodenough,\cite{ref11} when the cation-occupied octahedral share a common corner, there can be no direct overlap of neighboring cation orbitals and therefore there is no cation-cation interactions. 

What is the origin of $J_{1}$? The point C$_{1}$ is the center of the neighboring octahedron. The direction of C$_{1}$O$_{1}$ is the $z$ axis of this octahedron. A $p$ orbital of O$_{1}$ is expected to be directed toward C$_{1}$ ($p_{\sigma}$ orbital) so as to overlap the $d(3z^{2}-r^{2})$ orbital of the point C$_{1}$, where the $p$ orbital ($p_{x}$, $p_{y}$, $p_{z}$ states) is called as $p_{\sigma}$ orbital when the principal axis of the $p$ orbital coincides with the direction of the bond. A partial covalent bond between the $d(3z^{2}-r^{2})$ orbital and the $p_{\sigma}$ orbital can be formed. Then the charge transfer occurs from the $p_{\sigma}$ orbital with the spin-up state $\mid\uparrow \rangle$ to the $\epsilon_{g}$ state. Consequently, the spin of Mn$^{4+}$ at the point C$_{1}$ is still in the spin-up state, while the resulting spin of O$^{2-}$ at the point O$_{1}$ is in the spin-down state. The remaining $p_{\sigma}$ state at the point O$_{1}$ is magnetically coupled with the $t_{2g}$ state of the Mn$^{4+}$ at the point B$_{1}$. When this coupling (denoted as $K_{0}$) is antiferromagnetic, then the suprexchange interaction between Mn$^{4+}$ at the point C$_{1}$ and Mn$^{4+}$ at the point B$_{1}$ is ferromagnetic. On the other hand, when this coupling $K_{0}$ is ferromagnetic, then the superexchange interaction between Mn$^{4+}$ at the point C$_{1}$ and Mn$^{4+}$ at the point B$_{1}$ is antiferromagnetic. Here we note that the angle $\alpha=\angle$C$_{1}$O$_{1}$B$_{1}$ is equal to $130.62^{\circ}$ for $\beta$-MnO$_{2}$, which is very different from $90^{\circ}$. If $\alpha = 90^{\circ}$, the $p_{\sigma}$ orbital on the bond O$_{1}$C$_{1}$ coincides with the $p_{\pi}$ orbital on the bond O$_{1}$B$_{1}$ (the $y$ axis), where the $p$ orbital ($p_{x}$, $p_{y}$, $p_{z}$ states) is called as $p_{\pi}$ orbital when the principal axis of the $p$ orbital is perpendicular to the direction of the bond. The $p_{\pi}$ orbital is coupled with the $t_{2g}$ state of the Mn$^{4+}$ at the point B$_{1}$, since the $\epsilon_{g}$ state is empty. According to the Goodenough-Kamamori-Anderson rule,\cite{ref11,ref12,ref14,ref15} the interaction $K_{0}$ is antiferromagnetic, since the $p_{\pi}$ orbital is not orthogonal to the $t_{2g}$ orbital. If $\alpha = 180^{\circ}$, the $p_{\sigma}$ orbital is coupled with the $t_{2g}$ orbital of the Mn$^{4+}$ at the point B$_{1}$ the interaction $K_{0}$ is ferromagnetic, since the $p_{\sigma}$ orbital is orthogonal to the $t_{2g}$ orbital.

The sign of $K_{0}$ is dependent on the value of $\alpha$. There may be a critical angle $\alpha_{c}$. $K_{0}$ is ferromagnetic for $\alpha >\alpha_{c}$ and $K_{0}$ is antiferromagnetic for $\alpha <\alpha_{c}$. Experimentally, the superexchange interaction $J_{1}$ is antiferromagnetic for $\beta$-MnO$_{2}$, which means that $K_{0}$ is ferromagnetic. The critical angle $\alpha_{c}$ is lower than $130.62^{\circ}$. 

The direct cation-cation exchange interaction ($J_{2}$) is expected to be stronger than the superexchange interactions $J_{1}$. The competition between $J_{1}$ and $J_{2}$ can lead to a complicated compromise magnetic order; $J_{2}/J_{1} = 1.68469$ for $\beta$-MnO$_{2}$. Note that the discussion of Osmond\cite{ref16} on the nature of $J_{2}$ may be inappropriate. 

\subsection{\label{disC}Origin of superexchange interaction $J_{3}$} 
We consider the interaction between Mn$^{4+}$ at the point A$_{1}$ and the Mn$^{4+}$ at the point A$_{2}$, as shown in Figs.~\ref{fig20}(a) and (b), where the distance A$_{1}$A$_{2}$ is $4.396 \AA$. According to Goodenough,\cite{ref11,ref12} this interaction ($J_{3}$) is a superexchange one (cation-anion-cation) since the cation-occupied octahedra share a common corner (point O$_{4}$). The angle $\angle$A$_{1}$O$_{4}$A$_{2}$ is equal to $\alpha = 111.60^{\circ}$, the distance O$_{4}$A$_{2}$ is $3.343 \AA$, and the distance O$_{4}$A$_{1}$ is $1.878 \AA$. This interaction is experimentally antiferromagnetic for $\beta$-MnO$_{2}$; $p_{3}=J_{3}/J_{1} = 0.537$. This means that the critical angle $\alpha_{c}$ is between $90^{\circ}$ and $111.60^{\circ}$. 

\section{CONCLUSION} 
We have studied the phase diagram of ($J_{2}$ vs $J_{3}$) with $J_{1} = -1$ in the rutile type $\beta$-MnO$_{2}$ by using the equi-energy contour plot. The distribution of the magnetic Bragg peaks can be clearly visualized. The magnetic phase diagram consists of the multricritical point (the intersection $J_{2}J_{3}=J_{1}^{2}$ and $J_{2}+J_{3} = 2J_{1}$), the helical order along the $c$ axis, the $(h=1/2,k=0,l=1/2)$ phase, the helical order along the $a$ axis, and the phase $(h= 0,k=0,l=1)$. The phase transition is of the first order between the $(h=1/2,k=0,l=1/2)$ phase and the helical order along the $c$ axis, and between the $(h=1/2,k=0,l=1/2)$ phase and the helical order along the $a$ axis. The phase transition is of the second order between the phase $(h=0,k=0,l=1)$ and the helical order along the $c$ axis, and between the phase $(h=0,k=0,l=1)$ and the helical order along the $a$ axis.

\begin{acknowledgments}
We are grateful to Prof. H. Sato for useful discussions on the itinerant nature of $e_{g}$ electrons in $\beta$-MnO$_{2}$. 
\end{acknowledgments}


\begin{references}
\bibitem{ref01} R.A. Erickson, Phys. Rev. \textbf{90}, 779 (1953). 
\bibitem{ref02} A. Yoshimori, J. Phys. Soc. Jpn. \textbf{14}, 807 (1959).
\bibitem{ref03} N. Ohama and Y. Hamaguchi, J. Phys. Soc. Jpn. \textbf{30}, 1311 (1971). 
\bibitem{ref04} M. Regulski, R.Przenioslo, I. Sosnowska, and J.-U. Hoffmann, Phys. Rev. B \textbf{68}, 172401 (2003).
\bibitem{ref05} M. Regulski, R.Przenioslo, I. Sosnowska, and J.-U. Hoffmann, J. Phys. Soc. Jpn.\textbf{73}, 3444 (2004).
\bibitem{ref06} H. Sato, K. Wakiya, T. Enoki, T. Kiyama, Y. Wakabayashi, H. Nakano, and Y. Murakami, J. Phys. Soc. Jpn. \textbf{70}, 37 (2001).
\bibitem{ref07} H. Sato, Y. Kawamura, T. Ogawa, Y. Murakami, H. Ohsumi, M. Mizumaki, and N. Ikeda, Physica B \textbf{329 -- 333}, 757 (2003). 
\bibitem{ref08} H. Sato, T. Enoki, M. Isobe, and Y. Ueda, Phys. Rev. B \textbf{61}, 3563 (2000). 
\bibitem{ref09} T. Nagamiya, \textit{Solid State Physics}, edited by F. Seiz, D. Turnbull, and H. Ehrenreich, \textbf{20}, 305 (Academic Press, New York,1967). 
\bibitem{ref10} H. Bizette and B. Tsai, Colloque sur la polarization de la matere cited by Lidiard (Paris, C.N.R.S., 1949) 164, Reports on Progr. in Phys. \textbf{17}, 201 (1954). 
\bibitem{ref11} J.B. Goodenough, Phys. Rev.\textbf{117}, 1142 (1960). 
\bibitem{ref12} J.B. Goodenough, \textit{Magnetism and the Chemical Bond} (John-Wiley \& Sons, New York, 1963).
\bibitem{ref14} J. Kanamori, J. Phys. Chem. Solid \textbf{10}, 87 (1959).
\bibitem{ref15} P.W. Anderson, \textit{Solid State Physics}, edited by F. Seiz and D. Turnbull (Academic Press, New York and London, 1963) \textbf{14}, 99. 
\bibitem{ref16} W.P. Osmond, Proc. Phys. Soc. \textbf{87}, 335 (1966).
\end{references}
\end{document}